\documentclass[reprint,superscriptaddress,amsmath,amssymb,aps]{revtex4-1}
\usepackage{graphicx}
\usepackage[caption=false]{subfig}
\usepackage{multirow}
\usepackage{dcolumn}
\usepackage{bm}
\usepackage{hyperref}
\hypersetup{colorlinks=True,
            linkcolor=blue,
           anchorcolor = blue,
            citecolor = blue,
            filecolor = blue,
            urlcolor = blue }       
\DeclareMathOperator{\sech}{sech}
\begin{document}
\title{Coherent population transfer with polariton states in circuit QED }
\author{Madan Mohan Mahana}
\affiliation{Department of Physics, Indian Institute of Technology Guwahati, Guwahati 781039, Assam, India}
\author{Sankar Davuluri}
\affiliation{Department of Physics, Birla Institute of Technology and Science Pilani, Hyderabad Campus, Hyderabad 500078, India}
\author{Tarak Nath Dey}
\email{tarak.dey@iitg.ac.in}
\affiliation{Department of Physics, Indian Institute of Technology Guwahati, Guwahati 781039, Assam, India}

\begin{abstract}
    This article proposes a new method to increase the efficiency of stimulated Raman adiabatic passage (STIRAP) in superconducting circuits using a shortcut to the adiabaticity (STA) method. The STA speeds up the adiabatic process before decoherence has a significant effect, thus leading to increased efficiency. This method achieves fast, high-fidelity coherent population transfer, known as super-adiabatic STIRAP (saSTIRAP), in a dressed state-engineered $\Lambda$ system with polariton states in circuit QED. 
\end{abstract}
\maketitle
\section{\label{sec:level1}INTRODUCTION}
Superconducting quantum circuits (SQCs) are considered the leading solid-state quantum platforms owing to their extensive applications in quantum information processing and microwave quantum optics \cite{You2011, Blais2020}. The SQCs with state-of-the-art dilution refrigerators have minimal decoherence rates at milli Kelvin temperatures, which is the key success for realization quantum experiments.  The qubits of Josephson-junction-based SQCs at cryogenic temperature are the heart of quantum processors. The striking achievement on losses, radiation confinement, scalability and robustness of circuit quantum electrodynamics (circuit QED) has emerged as a new field of research that studies the interaction of superconducting artificial atoms (SAAs) with microwave photons in SQCs \cite{RevModPhys.93.025005}. The high coherence time and tunability of qubits/atoms are paramount for quantum experiments. Natural atoms have very high coherence times, but precisely controlling their quantum parameters is extremely difficult. The artificially engineered analogues of atoms in solid-state quantum platforms enable us to control the coherence time and tunability to our liking \cite{GU20171}. SAAs like Transmon \cite{PhysRevA.76.042319} and Fluxonium \cite{doi:10.1126/science.1175552} are the best among the currently available artificial atoms in terms of coherence time and tunability.

In quantum optics, counter-intuitive phenomena such as Electromagnetically induced transparency (EIT) \cite{PhysRevLett.64.1107}, Autler-Townes splitting (ATS) \cite{PhysRev.100.703}, and coherent population trapping (CPT) \cite{Alzetta1976}  has a significant role in the precise control of the optical property of a medium. These atomic coherence-based experiments demand an atomic configuration with a larger atomic coherence lifetime. A three-level $\Lambda$ system containing two longer-lived lower-level ground states can be fulfilled the said criterion. Hence, the three-level $\Lambda$ systems are more suitable for realizing EIT, ATS and CPT than ladder ($\Xi$) and V-type configurations. The EIT is an essential mechanism to suppress a weak probe field's absorption in the presence of a strong control field. The strong control field opens up an extra excitation pathway, which destructively interferes with the probe-assisted pathway. As a result, it creates a narrow transparency window that appears in the probe field's absorption spectrum \cite{RevModPhys.77.633}.  ATS is similarly associated with a dip in the spectral line of the probe field, resulting from splitting the spectral line by a resonant strong control field \cite{Cohen-Tannoudji1996}. The three-level quantum systems in SQCs have been used to demonstrate EIT \cite{PhysRevLett.104.193601,Novikov2016}, ATS \cite{PhysRevLett.102.243602,PhysRevB.88.060503,Suri_2013}, and CPT \cite{PhysRevLett.104.163601}.  

STIRAP is another example of a counter-intuitive phenomenon where robust population transfer between two nondegenerate metastable levels
is possible without loss of generality \cite{GAUBATZ1988463}. In STIRAP, a suitable choice of two time-dependent coherent pulses that are coupled to two arms of a three-level $\Lambda$ system allows a complete population transfer from the ground state to the target meta-stable state without populating the intermediate excited state. STIRAP has been experimentally realized in many quantum optical systems, including SQCs \cite{Kumar2016, Xu2016}.  Many studies have been devoted to the process of STIRAP systems with SQCs \cite{SIEWERT2006435,PhysRevB.87.214515,PhysRevB.91.224506,PhysRevA.93.051801,Zheng2022}. STIRAP-based population transfer has numerous applications in quantum optics and quantum information processing \cite{RevModPhys.89.015006}. 

The adiabatic processes are associated with slow change of controls, which leave some dynamical properties invariant. In the quantum regime, slow processes with long operational times are affected by decoherence, which produces losses and errors. The STA methods are well-established techniques to speed up the adiabatic protocols and achieve the same final results \cite{PhysRevLett.104.063002}. Counterdiabatic driving (CD) \cite{Berry_2009}, Lewis-Reisenfeld invariant (LRI) method \cite{PhysRevA.83.062116,PhysRevA.86.033405}, the dressed-state approach \cite{PhysRevLett.116.230503} are useful tools for STA techniques \cite{PhysRevA.94.063411} to speed up the adiabatic quantum protocols \cite{RevModPhys.91.045001}. Remarkably, STIRAP can be speed up by applying STA methods.  The successful application of CD protocol with STIRAP has been implemented in a three-level ladder-type superconducting qutrit \cite{doi:10.1126/sciadv.aau5999}. The super-adiabatic population transfer (saSTIRAP) from the ground state to the second excited state can be accomplished at a two-photon process.  However, two-photon detuning produces small ac-Stark shifts to all the energy levels hold drawback of the system. This issue can be resolved by dynamically modifying the phases of all applied drives.  The application of  the CD control field in a $\Lambda$ system, which drives the transition from the initial ground state to the target metastable state forms a closed loop $\Delta$ system. To the best of our knowledge, there has not been a theoretical investigation of saSTIRAP with a closed loop $\Lambda$ system in SQCs yet.

The experimental realization of $\Lambda$ systems with meta-stable states with SQCs has been elusive. The implementation of a dressed-state engineered impedance-matched $\Lambda$ system in the polariton basis has been investigated \cite{PhysRevLett.111.153601}. It opened up the avenue for theoretical and experimental demonstration of several quantum optical applications with $\Lambda$ system in SQCs \cite{PhysRevLett.113.063604,Inomata2016,PhysRevApplied.7.064006}. The implementation of EIT with an identical system has been theoretically proposed in \cite{PhysRevA.93.063827} and has been experimentally realized in polariton states generated with a rf-biased two-level system coupled to a resonator \cite{PhysRevLett.120.083602}. With all the dipole allowed transitions, a closed loop $\Delta$ configuration is possible which is rather impossible in natural atoms. We exploit this advantage to theoretically study the possibility of the experimental realization of STIRAP and saSTIRAP protocols with a driven circuit QED system. We use the doubly-dressed polariton states instead of the qutrit states \cite{doi:10.1126/sciadv.aau5999} used recently to study the coherent transfer of population in SQCs. The SAAs like flux qubits \cite{PhysRevLett.95.087001} are operated away from the sweet spot to break the parity-selection rule and form $\Delta$-type configurations. However, the driven circuit QED system suggested in this paper remedies this issue while still maintaining the coherence properties of the sweet spot.

The paper is organized as follows. In section \ref{sec:level2}, we describe the theoretical model of the Hamiltonian and the tunable transition rates of a $\Lambda$ system in circuit QED. We discuss the theoretical proposal for implementing CD protocol in section \ref{sec:level3}. Section \ref{sec:level4} thoroughly discusses the significant results. Finally, we conclude in section \ref{sec:level5}.
\section{\label{sec:level2}Theoretical Model}
This section deliberates the theoretical model for realizing a $\Lambda$ system in circuit QED. First, we describe the Hamiltonian of the model, then derive the expression for transition rates of the $\Lambda$ system.
\begin{figure}[t!]
    \centering
    \includegraphics[height=4.5cm,width=0.48\textwidth]{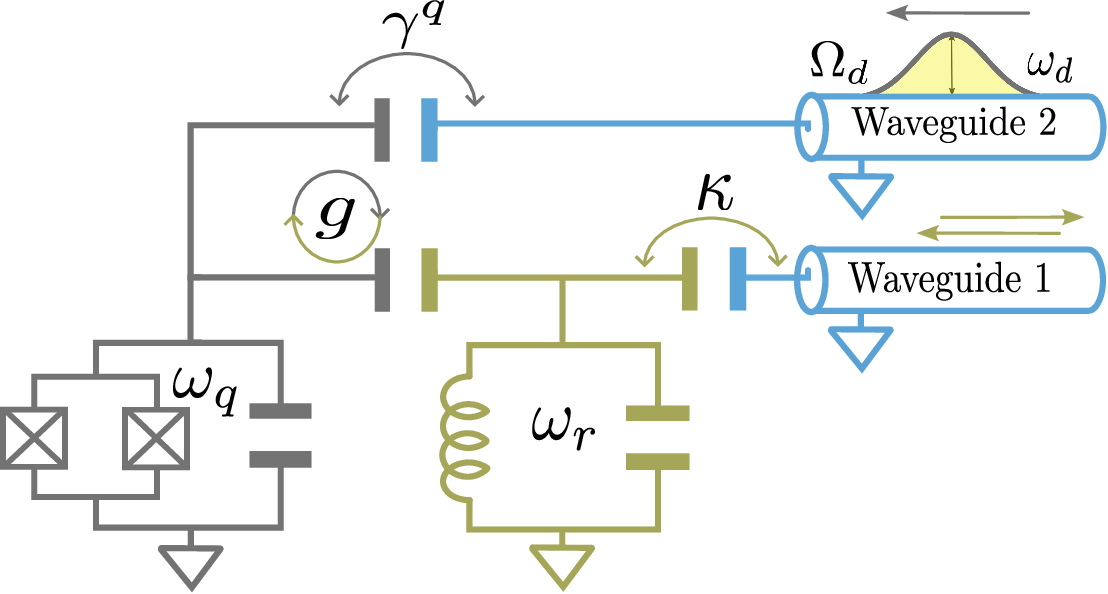}
    \caption{A schematic lumped-element circuit diagram of a driven circuit QED system. Here, a classical microwave field with coupling strength $\Omega_d$ drives a transmon with frequency $\omega_d$, further capacitively coupled to a single mode cavity with coupling strength $g$. Both the transmon and cavity are capacitively coupled to two independent semi-infinite waveguides.}
    \label{fig:fig1}
\end{figure}
\subsection{\label{sec:level2a}Model Hamiltonian}
The ability to precisely control a quantum system's population at various levels is challenging due to decay-induced population loss. The STIRAP is an indispensable tool for transferring the population to the desired levels and overcoming the decay-induced population loss. A counter-intuitive pair of laser pulses is used in the STIRAP. A pump pulse couples between an initial populated ground state with an excited state, whereas an advanced Stokes pulse couples an unoccupied metastable state and an excited state. The two lower-level metastable states coupling with the intermediate excited state by the coherent pulses form $\Lambda$ type configuration. In this level geometry, the efficiency and robustness of the population transfer are sensitive to the overlaps between the Stoke and the pump fields and the individual pulse area. The slow rate of population transfer in STIRAP is the obstacle to efficient population transfer and the reason behind the population loss. Hence, faster population transfer processes such as saSTIRAP can avoid inherent decay and decoherence limitations. This work explores speeding up STIRAP passage in circuit quantum electrodynamics by considering a closed loop $\Lambda$ system in which the ground and meta-stable states can be directly coupled. The scheme for a coupled transmon-cavity system is shown schematically in Fig. \ref{fig:fig1}. The cavity and the transmon are connected to a semi-infinite waveguide 1 and 2, respectively. From waveguide 2, a microwave field with coupling strength $\Omega_d$ drives a two-level transmon with frequency $\omega_d$, further capacitively coupled to a single mode cavity with coupling strength $g$. The total Hamiltonian of the system can be cast in the following form
\begin{align}\label{1}
  H_0&= \frac{\hbar}{2}\omega_q\sigma_z+\hbar\omega_r\left(a^\dag a+\frac{1}{2}\right)+\hbar g(a^\dag\sigma_-+a\sigma_+)\nonumber\\
   & +\hbar\Omega_d\left(\sigma_-e^{i\omega_dt}+\sigma_+e^{-i\omega_dt}\right),
\end{align}
where $\omega_q$ and $\omega_r$ stand for the transmon and the cavity frequencies, respectively. The annihilation and creation operator of the cavity are denoted by $a$ and $a^\dag$, whereas the atomic lowering and raising operators for transmon are $\sigma_-$ and  $\sigma_+$. The interaction strength and frequency of the classical microwave field are expressed by the parameters $\Omega_d$ and $\omega_d$, respectively. We eliminate the explicit time-dependent factors of the Hamiltonian by transforming the Hamiltonian into a rotating frame using a unitary operator $U=exp{(-i\omega_d(\sigma_z/2+a^\dag a)t)}$ and obtain an effective Hamiltonian 
\begin{align}\label{2}
    H_{RWA}= & \frac{\hbar}{2}\tilde{\omega}_q\sigma_z+\hbar\tilde{\omega}_r\left(a^\dag a+\frac{1}{2}\right)+\hbar g(a^\dag\sigma_-+a\sigma_+)\nonumber\\
     & +\hbar\Omega_d[\sigma_-+\sigma_+],
\end{align}
under rotating wave approximation. Here, $\tilde{\omega}_q=\omega_q-\omega_d$, $\tilde{\omega}_r=\omega_r-\omega_d$ and $\tilde{\Delta}=\tilde{\omega}_r-\tilde{\omega}_q$ is the cavity-transmon detuning.
The first three terms in the Hamiltonian can be identified as the celebrated Jaynes-cummings model. The last term represents the interaction between the external classical microwave drive field and the two-level transmon. The eigenstates of the Jaynes-Cummings Hamiltonian are known as the dressed states, which can be deliberate as
\begin{align}\label{44}
    |+,n\rangle & = \cos{\frac{\theta_n}{2}}|e,n\rangle+\sin{\frac{\theta_n}{2}}|g,n+1\rangle,\\
    |-,n\rangle & = -\sin{\frac{\theta_n}{2}}|e,n\rangle+\cos{\frac{\theta_n}{2}}|g,n+1\rangle,
\end{align}
where $\tan{\theta_n}=-2g\sqrt{n+1}/\tilde{\Delta}$. Here, $|e,n\rangle$ and $|g,n\rangle$ denotes that the qubit is in the excited state $|e\rangle$ and ground state $|g\rangle$, respectively, whereas the single-mode cavity is in the state $|n\rangle$. The corresponding eigenvalues of the dressed states are
\begin{equation}\label{45}
   E_{\pm,n}=\hbar\tilde{\omega}_r(n+1)\pm\frac{\hbar}{2}\sqrt{\tilde{\Delta}^2+4g^2(n+1)}.
\end{equation}
Further mixing of these dressed states in the dispersive regime ($g\ll\tilde{\Delta}$) by the external microwave field applied to drive the transmon gives doubly dressed polariton states. Polaritons are referred to as quasi-particles carrying elementary excitations of the light-matter interaction. These polariton states can be denoted by $|i\rangle$, $|j\rangle$ with the corresponding eigenenergies $\omega_i$, $\omega_j$ $(i,j=1,2,3,4,..)$. The polariton states can be engineered to obtain a nested four-level system consisting of the lowest four eigenstates of (\ref{2}) by restricting the driving field to satisfy the condition $\omega_q-3\chi<\omega_d<\omega_q-\chi$, where $\chi=g^2/\tilde{\Delta}$ denotes the dispersive frequency shift \cite{PhysRevLett.111.153601,PhysRevA.93.063827}. Under the so-called nesting regime, the levels $|1\rangle,|3\rangle$ (or $|4\rangle$), and $|2\rangle$ form a $\Lambda$ system configuration as shown below in Fig. (\ref{fig:fig2}).
\begin{figure}[h!]
    \centering
    \includegraphics[height=3cm, width=5cm]{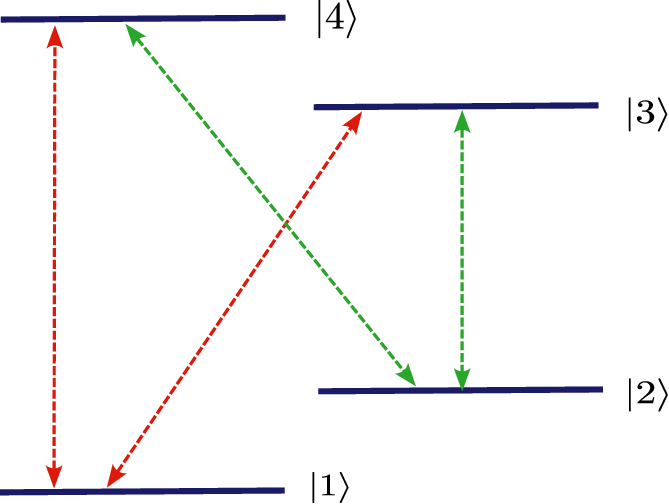}
    \caption{Energy-level diagram of the doubly-dressed polariton states in the driven circuit QED system.}
    \label{fig:fig2}
\end{figure}
\subsection{\label{sec:level2b}Transition rates}
Waveguides 1 and 2 are coupled to the cavity and transmon to apply drive and readout signals. We consider two waveguides as external environments, and the total Hamiltonian of the open quantum system can be written as
 \begin{equation}\label{3}
    H_{T}=H_S+H_E+H_I,
\end{equation}
where $H_S$, $H_E$, and $H_I$ denote Hamiltonians of the system, the environment, and the system-environment interaction, respectively. We consider Eq. (\ref{1}) as the Hamiltonian of the system $H_S$. The Hamiltonian of the environment $H_E$ is expressed as
\begin{equation}\label{4}
    H_E=\hbar\int \omega b^\dag(\omega)b(\omega) d\omega +\hbar\int \omega^\prime c^\dag(\omega^\prime)c(\omega^\prime)d\omega^\prime,
\end{equation}
where $b$ and $c$ denote the annihilation operator in waveguide 1 and waveguide 2, respectively, and $b^\dag$ and $c^\dag$ represent the corresponding creation operator. Finally, the system-environment interaction Hamiltonian is denoted by $H_I$ and can be defined as 
\begin{equation}\label{5}
\begin{aligned}
    H_I= & \hbar\left[\int d\omega K(\omega)b^\dag(\omega)a+H.C.\right]\\
    +&\hbar\left[\int d\omega^\prime\eta(\omega^\prime)c^\dag(\omega^\prime)\sigma_-+H.C.\right].
\end{aligned}
\end{equation}
The Hamiltonian $H_E$ and $H_I$ remain unchanged in the rotating frame. For simplicity, we can consider having flat spectra in the environments so that both $K(\omega)$ and $\eta(\omega^\prime)$ can be constants. Introducing the first Markov approximation, we can deliberate
\begin{align}\label{6}
    K(\omega)= & \sqrt{\frac{\kappa}{{2\pi}}},\\
    \eta(\omega^\prime)= & \sqrt{\frac{\gamma^q}{{2\pi}}},
\end{align}
where $\kappa$ and $\gamma^q$ denote the decay rate of the cavity and transmon, respectively, into waveguides 1 and  2. Let us write the operators $a$ and $\sigma_-$ in the polariton basis
\begin{align}\label{7}
    a= & \sum\limits_{ij}\langle i|a|j\rangle\sigma_{ij},\\
    \sigma_-= & \sum\limits_{ij}\langle i|\sigma_-|j\rangle\sigma_{ij},
\end{align}
where $|i\rangle$, $|j\rangle$ denote the polariton states and $\sigma_{ij}=|i\rangle\langle j|$. In the polariton basis, the Hamiltonian $H_I$ can be recast into the following form
\begin{equation}\label{8}
\begin{aligned}
    H_I= & \hbar\int d\omega\sum\limits_{ij}\left[\sqrt{\frac{\kappa_{ij}}{2\pi}}b^\dag(\omega)\sigma_{ij}+H.C.\right]\\
    & +\hbar\int d\omega^\prime\sum\limits_{ij}\left[\sqrt{\frac{\gamma^q_{ij}}{2\pi}}c^\dag(\omega^\prime)\sigma_{ij}+H.C.\right],
\end{aligned}
\end{equation}
where $\kappa_{ij}$ and $\gamma^q_{ij}$  are the radiative decay rates into waveguide 1 and waveguide 2, respectively, for the transition from polariton state $|i\rangle$ to $|j\rangle$. The transition rates $\kappa_{ij}$, $\gamma^q_{ij}$ are defined as
\begin{align}\label{9}
    \kappa_{ij}= & \kappa|\langle i|a^\dag|j\rangle|^2,\\
    \gamma^q_{ij}= & \gamma^q|\langle i|a^\dag|j\rangle|^2.
\end{align}
Hence, we can determine the total radiative decay rate $\gamma_{ij}$ for transition between polariton states $|i\rangle$ and $|j\rangle$ as follows
\begin{equation}\label{10}
    \gamma_{ij}=\kappa_{ij}+\gamma_{ij}^q=\kappa C^2_{ij}+
    \gamma^q Q^2_{ij},
\end{equation}
where the parameters $C_{ij}=|\langle i|a^\dag|j\rangle|$ and $Q_{ij}=|\langle i|\sigma_+|j\rangle|$ represent the transition matrix elements corresponding to external drives applied to the cavity and the qubit respectively.
\begin{table}[h!]
    \centering 
    \begin{ruledtabular}
    \begin{tabular}{c c c c}
    Parameter & value & Parameter & Value\\
    \hline
    $C_{31}$ & 0.77 & $\omega_{31}$ & 5101\\
    $C_{32}$ & 0.64 & $\omega_{32}$ & 5023\\
    $C_{21}$ & 0.08 & $\omega_{21}$ & 78\\
    $Q_{31}$ & 0.00 & $\gamma_{31}$ & 7.47\\
    $Q_{32}$ & 0.10 & $\gamma_{32}$ & 5.18\\
    $Q_{21}$ & 0.82 & $\gamma_{21}$ & 0.96
    \end{tabular}
    \end{ruledtabular}
    \caption{Numerically calculated values for the transition matrix elements ($C_{ij},Q_{ij}$), radiative transition rates ($\gamma_{ij}$), and the transition frequencies ($\omega_{ij}=\omega_i-\omega_j$) in the polariton basis. The units of $\gamma_{21}$, $\gamma_{32}$, $\gamma_{31}$, $\omega_{21}$, $\omega_{32}$ and $\omega_{31}$ are in $2\pi$ MHz units. The parameters $\omega_q/{2\pi}=5$ GHz, $\omega_r/{2\pi}=10$ GHz, $\omega_d/{2\pi}=4.9$ GHz, $g/{2\pi}=0.5$ GHZ, $\Omega_d/{2\pi}=30$ MHz, $\kappa/{2\pi}=3$ MHz and $\gamma^q/{2\pi}=0.2$ MHz, and the exact eigenstates of Hamiltonian (\ref{2}) are used for the numerical calculation of the above parameters.} 
    \label{tab:my_label}
\end{table}
The energies of the polariton states can be tuned by the frequency, $\omega_d$ and the Rabi frequency $\Omega_d$ of the classical microwave drive field applied to the transmon through waveguide 2. Thus, the decay rates $\gamma_{ij}$ can also be tuned by varying the above parameters. By assigning constant values to these parameters, one can design a $\Lambda$ system with fixed energy levels and transition rates. The list of numerically computed values of the relevant parameters for our $\Lambda$ system is tabulated in the table \ref{tab:my_label}.
\section{\label{sec:level3}Counteradiabatic driving}
The STIRAP process can be implemented with the three-level $\Lambda$ system described in section \ref{sec:level2}. The matrix representation of the STIRAP Hamiltonian under rotating wave approximation is
\begin{equation}\label{11}
    H(t)=\frac{\hbar}{2}
    \begin{pmatrix}
    0 & \Omega_p(t) & 0\\
    \Omega_p(t) & 2\Delta & \Omega_s(t)\\
    0 & \Omega_s(t) & 2\delta
    \end{pmatrix},
\end{equation}
where $\Omega_p(t)$ and $\Omega_s(t)$ denote the coupling strength of the time-dependent pump and Stokes field for $|3\rangle\leftrightarrow|1\rangle$ and $|3\rangle\leftrightarrow|2\rangle$ atomic transitions with frequencies $\omega_p$ and $\omega_s$, respectively. The parameter $\Delta$ and $\delta$ denotes the one-photon detuning $\Delta=(\omega_{31}-\omega_p)$ and two-photon detuning $\delta=(\omega_{31}-\omega_p)-(\omega_{32}-\omega_s)$, respectively.  Here, the energy levels of the $\Lambda$ system satisfy $E_1<E_2<E_3$. The one-photon detuning $\Delta$ differs from the cavity-transmon detuning $\tilde{\Delta}$, discussed in the last section. For a perfectly resonant STIRAP process, {\it i.e.}, $\Delta=\delta=0$, the instantaneous eigenvalues of the above Hamiltonian are $E_0=0$ and $E_{\pm}=\pm\hbar\Omega_0(t)/2$ with $\Omega_0(t)=\sqrt{\Omega^2_p(t)+\Omega^2_s(t)}$. The corresponding instantaneous eigenstates are written as
\begin{equation}\label{12}
    |n_0(t)\rangle=
    \begin{pmatrix}
        \cos{\theta(t)}\\
        0\\
        -\sin{\theta(t)}
    \end{pmatrix},
    |n_{\pm}(t)\rangle=\frac{1}{\sqrt{2}}
    \begin{pmatrix}
        \sin{\theta(t)}\\
        \pm 1\\
        \cos{\theta(t)}
    \end{pmatrix},
\end{equation}
where $\tan{\theta(t)}=\Omega_p(t)/\Omega_s(t)$. Perfect adiabatic population transfer from state $|1\rangle$ to $|2\rangle$ can be achieved by following the dark state $|n_0(t)\rangle$, under the local adiabatic condition $|\dot{\theta}|\ll|\Omega_0|$ need to be fulfilled \cite{PhysRevA.40.6741}.

\begin{figure}[t!]
    \centering
    \includegraphics[height=3cm,width=5cm]{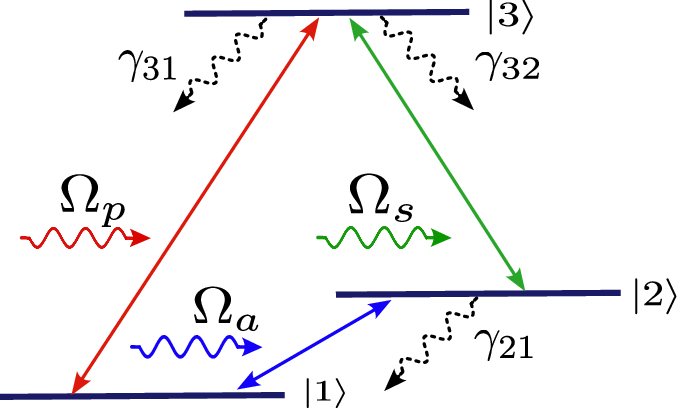}
    \caption{The schematic diagram of a three-level $\Delta$ system driven by three external classical fields $\Omega_p$, $\Omega_s$ and $\Omega_a$ respectively.}
    \label{fig:fig3}
\end{figure}

To speed up the STIRAP protocol, one can apply an additional coupling field driving the $|2\rangle\leftrightarrow|1\rangle$ transition \cite{Berry_2009, PhysRevLett.105.123003, PhysRevA.89.033419}. The additional drive is termed a counter-diabatic drive (CD) or transition-less quantum drive (TQD), and it can be expressed as
\begin{equation}\label{13}
    H^{CD}(t)=i\hbar\sum\limits_n[|\partial_t n(t)\rangle\langle n(t)|-\langle n(t)|\partial_t n(t)\rangle|n(t)\rangle\langle n(t)|].
\end{equation}
We derive the Hamiltonian $H^{CD}(t)$ using the adiabatic basis states $|n(t)\rangle=(|n_0(t)\rangle,|n_{\pm}(t)\rangle)$, which reads
\begin{equation}\label{14}
    H^{CD}(t)=\frac{\hbar}{2}
    \begin{pmatrix}
        0 & 0 & i\Omega_a(t)\\
        0 & 0 & 0\\
        -i\Omega_a(t) & 0 & 0
    \end{pmatrix},
\end{equation}
where $\Omega_a(t)=2\dot{\theta}(t)$ and the overdot denotes the first derivative with respect to time. We assume that the external drives are applied to the dressed state-engineered $\Lambda$ system with polariton states by driving the cavity and the transmon qubit. The Hamiltonian representing the interaction between the transmon and external drive fields is thus given by
\begin{equation}\label{15}
    H_{d}=\frac{\hbar}{2}(A_pa^\dag e^{-i\omega_pt}+A_sa^\dag e^{-i\omega_st}+A_a\sigma_+ e^{-i\omega_at}+H.C.),
\end{equation}
where the pump field and Stokes field with frequenciey $ \omega_p$ and $\omega_s$ respectively are coupled to the cavity with coupling strength $A_p$ and $A_s$. The additional drive field {\it i.e.,} counterdiabatic drive with frequency $\omega_a$ is coupled to the tansmon with coupling strength $A_a$. Considering the pump field, Stokes field, and the conterdiabatic drive field are driving the $|3\rangle\leftrightarrow|1\rangle$, $|3\rangle\leftrightarrow|2\rangle$ and $|2\rangle\leftrightarrow|1\rangle$ transition in the polariton basis respectively, we define the amplitudes of the respective Rabi frequencies of the external drive fields in the polariton basis as
\begin{equation}\label{16}
    \Omega_p\approx  A_pC_{31},~
    \Omega_s\approx  A_sC_{32},~
    \Omega_a\approx  A_aQ_{21},
\end{equation}
where the parameters $C_{31}$, $C_{32}$ and $Q_{21}$ are the transition matrix elements already defined in sec. \ref{sec:level2}. Here, $\Omega_p,\Omega_s$, and $\Omega_a$ denote the Rabi frequencies of the external drives coupled to the $\Lambda$ system in the polariton basis as shown in Fig. (\ref{fig:fig3}). Here, the Gaussian envelopes of the pump and Stokes fields are considered and stated as
\begin{align}\label{20}
    \Omega_p(t)= & \Omega_p e^{-\frac{t^2}{2\sigma^2}},\\
    \Omega_s(t)= & \Omega_s e^{-\frac{(t-t_s)^2}{2\sigma^2}}.
\end{align}
Using Eqs. (\ref{11})-(\ref{14}), one can obtain
\begin{equation}\label{21}
    \Omega_a(t)= -\frac{t_s}{\sigma^2}\sech{\left[-\frac{t_s}{\sigma^2}{\left(t-\frac{t_s}{2}\right)}\right]},
\end{equation}
with assumption of $\Omega_p=\Omega_s$ for brevity. Eq. (\ref{21}) shows that the counteradiabatic drive should have a Rabi frequency $\Omega_a=-t_s/\sigma^2$ with a sec-hyperbolic shape for the above pump and stokes fields.
Thus, we can write the total Hamiltonian of the system under rotating wave approximation as
\begin{equation}\label{18}
    \tilde{H}(t)=\frac{\hbar}{2}[\Omega_p(t) |3\rangle\langle 1|+\Omega_s(t) |3\rangle\langle 2|-i\Omega_a(t) |2\rangle\langle 1|+H.C.],
\end{equation}
where we consider the resonant drive conditions ($\delta=\Delta=0$), $\omega_p=\omega_{31}$, $\omega_s=\omega_{32}$ and $\omega_a=\omega_{21}$. In order to solve the time evolution of the system we adopt the Lindblad master equation \cite{inbook,RevModPhys.82.1155}
\begin{equation}\label{19}
    \dot{\rho}=\frac{1}{i\hbar}[\tilde{H},\rho] + \sum\limits_{j=1}^{3}\mathcal{L}(\mathcal{O}_j)\rho,
\end{equation}
where $\mathcal{L}(\mathcal{O}_j)\rho = (2\mathcal{O}_j\rho\mathcal{O}_j^\dagger-\rho\mathcal{O}_j^\dagger\mathcal{O}_j-\mathcal{O}_j^\dagger\mathcal{O}_j\rho)/2$. Here, the operators $\mathcal{O}_j$ denote the jump operators given by $\mathcal{O}_1 = \sqrt{\gamma_{31}}|1\rangle\langle 3|$, $\mathcal{O}_2 = \sqrt{\gamma_{32}}|2\rangle\langle 3|$ and $\mathcal{O}_3 = \sqrt{\gamma_{21}}|1\rangle\langle 2|$. We substitute the drive fields given in Eqs. (\ref{20}-\ref{21}) in Eq. (\ref{18}) and solve the time evolution of the system using the Lindblad master equation given in Eq. (\ref{19}). The well-established $mesolver$ routine in Qutip \cite{JOHANSSON20121760,JOHANSSON20131234} is used for solving the time-dependent Lindblad master equations.  The numerical results are discussed in the following section.

\begin{figure}[t!]
    \centering
    \includegraphics[width=0.48\textwidth]{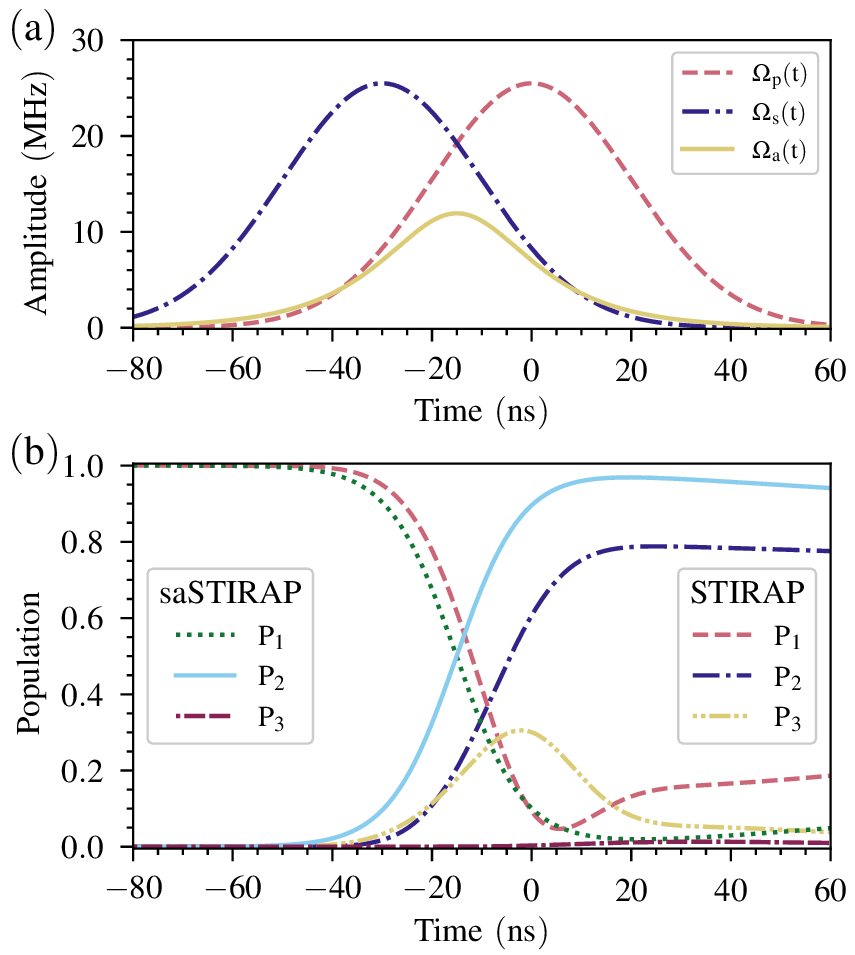}
    \caption{(a) The pulse-sequence of the three external classical drive fields $\Omega_p(t)$, $\Omega_s(t)$ and $\Omega_a(t)$.  (b) The time evolution of populations $\mathrm{P_1}$, $\mathrm{P_2}$ and $\mathrm{P_3}$ during STIRAP ($\Omega_a=0$) and saSTRIAP protocols. The parameters used for the simulation are $\Omega_p/{2\pi}=25.5$ MHz, $\Omega_s/{2\pi}=25.5$ MHz, $t_s=-30$ ns, $\sigma=20$ ns, and all the parameters given in table \ref{tab:my_label}.}
    \label{fig:fig4}
\end{figure}

\section{\label{sec:level4}Results and discussion}
We split this section into three parts describing the significant findings of this work. The Sec. \ref{sec:level4i} highlights the dynamics of coherent population transfer in our system. The sensitivity of the coherent transfer protocols to variations in the parameters is discussed in Sec. \ref{sec:level4ii}. We quantitatively compare the efficiencies of coherent population transfer protocols by numerically computing the fidelity in Sec. \ref{sec:level4iii}.

\subsection{\label{sec:level4i}Coherent population transfer}
We investigate the population dynamics in each energy level of the $\Lambda$ system described in section \ref{sec:level3}.  Fig. \ref{fig:fig4}(a) shows the pulse sequence of three external drive fields applied to the polariton state-$\Lambda$ system by driving the cavity mode. The Lindblad master equations for the STIRAP and the saSTIRAP protocols for the $\Lambda$ system are numerically solved to study the population dynamics. A counterdiabatic drive is applied to the $\Lambda$ system to realize the saSTIRAP by coupling $|3\rangle\leftrightarrow|1\rangle$ transition. Fig. \ref{fig:fig4}(b) substantiates that one can achieve faster coherent population transfer from level $|1\rangle$ to $ |2\rangle$ by applying the CD protocol as compared to the STIRAP in the $\Lambda$ system in the polariton basis. The populations $P_1,P_2$, and $P_3$ denote the populations in polariton states $|1\rangle,|2\rangle$, and $|3\rangle$, which are simply the density matrix elements $\rho_{11},\rho_{22}$, and $\rho_{33}$ respectively in the polariton basis. The numerical results proclaim that up to $78.81\%$ population can be transferred from the ground state to the meta-stable state by the STIRAP protocol with the used parameters. Furthermore, one can achieve a much higher efficiency up to $96.90\%$ population transfer with the saSTIRAP  protocol. Moreover, the efficiency of these protocols is also dependent on other important parameters such as the pulse amplitudes and the pulse widths of the external drive fields, normalized pulse separation, etc. that we elaborate on in the next section.

\subsection{\label{sec:level4ii}Sensitivity to parameters}

\begin{figure}[b!]
    \centering
    \includegraphics[height=7cm,width=0.48\textwidth]{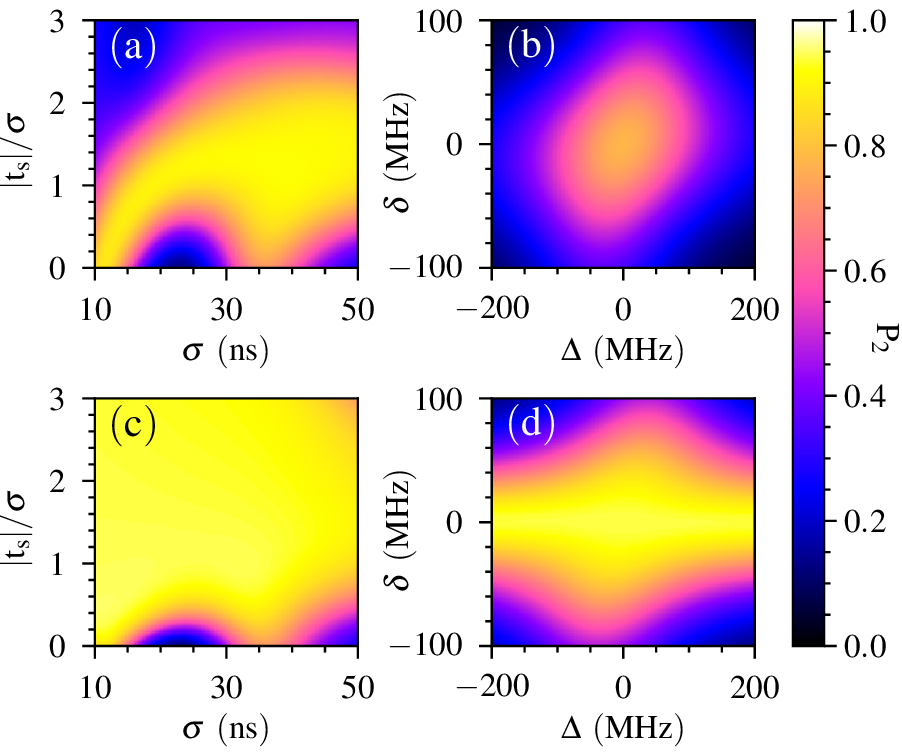}
    \caption{The efficiency of STIRAP (saSTIRAP) protocol in the parameter space of (a) ((c)) the pulse width ($\sigma$) and the normalized pulse separation ($|t_s|/\sigma$), (b) ((d)) the one-photon detuning ($\Delta$) and the two-photon detuning ($\delta$). The parameters used for the numerical simulation are $t_s=-30$ ns, $\sigma=20$ ns, and the parameters used in Fig.\ref{fig:fig4}.}
    \label{fig:fig5}
\end{figure}

This section benchmarks the efficiency of the STIRAP and saSTIRAP protocols with the variation in the parameters used for numerical analysis. At the end of these protocols, we quantify the population transfer efficiency by the final population in the state $|2\rangle$. Figs. \ref{fig:fig5}(a) and \ref{fig:fig5}(c) show how the efficiency varies in the parameter space of the pulse width $\sigma$ and the normalized pulse separation $|t_s|/\sigma$  in the STIRAP and the saSTIRAP protocols respectively. The parameters $\sigma=20$ ns, $t_s=-30$ ns for $\Omega_p/{2\pi}=\Omega_s/{2\pi}=25.5$ MHz lie in that highly efficient bright yellow regions in both the figures. The sensitivity of the STIRAP and saSTIRAP protocols to  are presented in Figs. \ref{fig:fig5}(b) and \ref{fig:fig5}(d) respectively. Fig. \ref{fig:fig5}(b) shows that the resonant driving condition ($\Delta=\delta=0$) is ideal for highly efficient coherent population transfer in the STIRAP protocol. Fig. \ref{fig:fig5}(d) shows the saSTIRAP protocol is more robust against one-photon detuning $\Delta$ than the two-photon detuning $\delta$. These figures indicate that the saSTIRAP protocol is more efficient and robust than the STIRAP protocol for a $\Lambda$ system in circuit QED.

\begin{figure}[b!]
    \centering
    \includegraphics[height=10cm,width=0.48\textwidth]{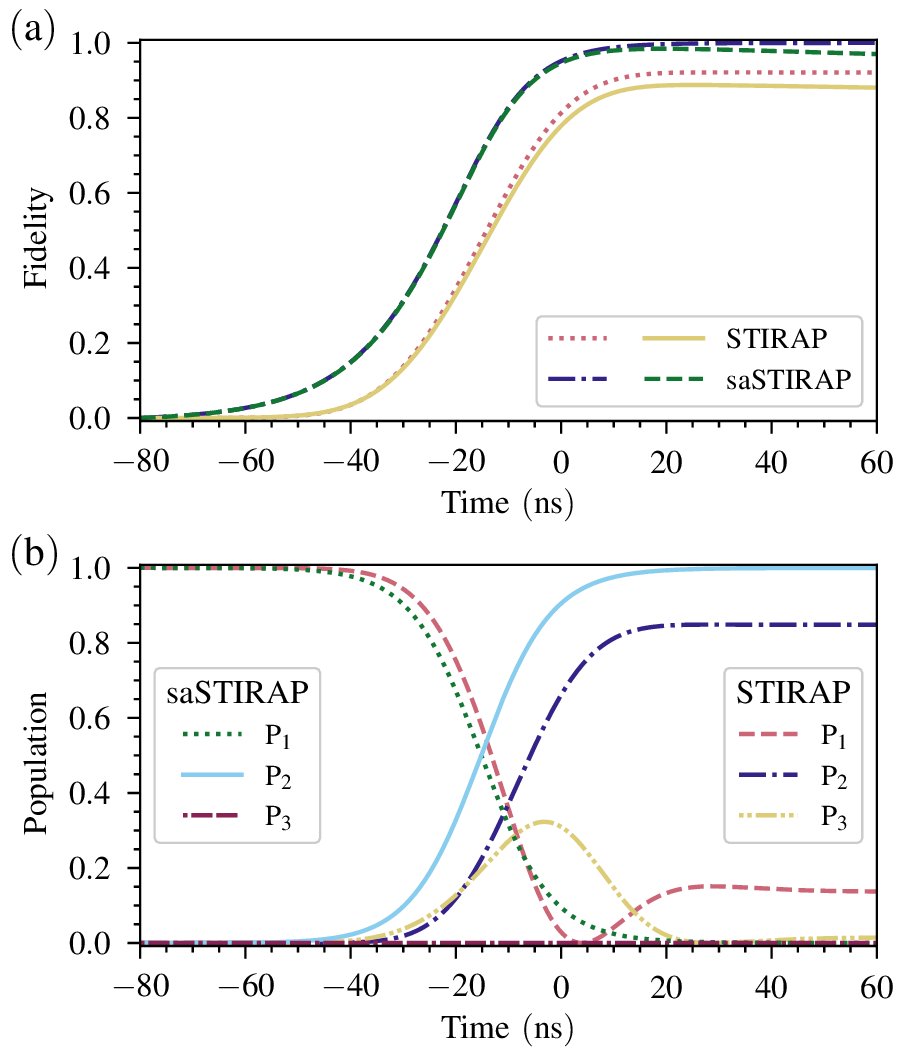}
    \caption{(a) Fidelity of the STIRAP (saSTIRAP) protocol with radiative transitions is shown with the solid-yellow (dashed-green) line and the dotted-red (dash-dotted-blue) line shows the fidelity with the radiative transition rates ($\gamma_{21},\gamma_{32}$, and $\gamma_{31}$) set to zero. (b) The time evolution of the populations in STIRAP and saSTIRAP protocols with the radiative transition rates ($\gamma_{21},\gamma_{32}$, and $\gamma_{31}$) set to zero. All the parameters used in Fig.\ref{fig:fig5} are used for the numerical simulation.}
    \label{fig:fig6}
\end{figure}

\subsection{\label{sec:level4iii}Fidelity}
In quantum information theory and quantum mechanics, the quantitative measure of the closeness of a quantum state at a final time `$t_f$' to the ideal target state is given by fidelity ($\mathcal{F}$) \cite{nielsen_chuang_2010}, and is defined as
\begin{equation}\label{46}
    \mathcal{F}(\rho_f,\rho_t)=\left(Tr\sqrt{\sqrt{\rho_t}\rho_f\sqrt{\rho_t}}\right)^2,
\end{equation}
where the density matrix operators $\rho_f$ and $\rho_t$ describe the quantum state of the system at time `$t_f$' and the ideal target state (here, $\rho_{22}$). 

In Fig. \ref{fig:fig6}(a), we have shown the fidelity between the final states of the time-evolution and the target state $|2\rangle$. One can observe that the fidelity of the saSTIRAP protocol is significantly higher than that of STIRAP over the operation time of the protocols.

Our numerical calculation suggests that the maximum fidelity for the saSTIRAP protocol is $98.44\%$, much higher than the maximum fidelity of $88.77\%$ for STIRAP, as shown in a solid black curve and long dashed curve. The dashed and dashed-dotted-lined plots in the above figure show the fidelity of both protocols in the absence of the radiative transition rates. We can observe a significant increment in the fidelity for each protocol without the radiative decay of the polariton states. Fig. \ref{fig:fig6}(b) shows the time evolution of the populations for each protocol without the radiative transitions, indicating higher efficiency of the population transfer. The numerical simulations suggest that a maximum of up to $84.88\%$ population can be transferred with $92.13\%$ fidelity with the STIRAP protocol without the radiative decay of the polariton states. The saSTIRAP protocol can transfer a maximum of up to $99.99\%$ population with $99.99\%$ fidelity with zero radiative transitions. Thus, we can realize highly efficient, robust, and high-fidelity coherent population transfer by increasing the coherence time of the transmon qubit and superconducting microwave resonator, reducing the radiative transitions in the polariton basis.

One can also implement the other STA techniques to speed up the transfer protocol with such systems in circuit QED and compare the efficiency and fidelity. 
\section{\label{sec:level5}Conclusion}
 In conclusion, we studied the application of STA to STIRAP using a dressed state-engineered system in circuit QED to achieve fast and high-fidelity coherent population transfer, known as saSTIRAP. An experimental realization of our theoretical proposal may use the currently available SQCs technologies. We further showed that the saSTIRAP technique leads to quantum state transfer with better fidelity than the STIRAP. The experimental realization of STIRAP and saSTIRAP in SQCs can find valuable applications in designing fast, high-fidelity quantum gates for efficient quantum computing and quantum information processing \cite{PhysRevA.77.022306,PhysRevA.76.062321}.

\bibliography{ref}
\end{document}